\documentclass{IEEEtran}
\usepackage{cite}
\usepackage{amsmath,amssymb,amsfonts}
\usepackage{graphicx}
\usepackage{textcomp,nicefrac}
\usepackage{lineno}
\usepackage{cleveref}

\def\BibTeX{{\rm B\kern-.05em{\sc i\kern-.025em b}\kern-.08em
T\kern-.1667em\lower.7ex\hbox{E}\kern-.125emX}}
\begin{document}
\title{Development of a 10 mol\% Rubidium-doped CsI Crystal for $^{87}$Rb Beta-Spectroscopy and Sterile Neutrino Searches}
\author{W.~K.~Kim, K.~W.~Kim, L.~T.~Truc, H.~S.~Lee, H.~J.~Kim, and Y.~D.~Kim
\thanks{This work was supported by the Institute for Basic Science (IBS), under project code IBS-R016-A1, Republic of Korea; The National Research Foundation of Korea (NRF), Grant funded by the
Korean government (MSIP) and Ministry of Science and Technology
(MEST), RS-2025-25460489.}
\thanks{W.~K.~Kim is with the IBS School, University of Science and Technology (UST), Daejeon 34113, Republic of Korea and Center for Underground Physics, Institute for Basic Science (IBS), Daejeon 34126, Republic of Korea.}
\thanks{L.~T.~Truc is with the Department of Physics, Kyungpook National University, Daegu 41566, Republic of Korea.}
\thanks{K.~W.~Kim is with the Center for Underground Physics, Institute for Basic Science (IBS), Daejeon 34126, Republic of Korea.}
\thanks{H.~S.~Lee and Y.~D.~Kim are with the Center for Underground Physics, Institute for Basic Science (IBS), Daejeon 34126, Republic of Korea and IBS School, University of Science and Technology (UST), Daejeon 34113, Republic of Korea.}
\thanks{H.~J.~Kim is with the Center for High Energy Physics, Kyungpook National University, Daegu 41566, Republic of Korea.}
\thanks{corresponding author : K.~W.~Kim (email: kwkim@ibs.re.kr) and W.~K.~Kim (email: wonkyung@ibs.re.kr)}
}
\maketitle

\begin{abstract}
The third-forbidden non-unique beta-decay of $^{87}$Rb to $^{87}$Sr (Q$_\beta = 282.275(6)$\,keV) has long served as an important benchmark for understanding forbidden beta-decay.
To investigate this, we have developed a novel CsI scintillator with a 10\,mol\% Rb concentration using the Bridgman method. The incorporated $^{87}$Rb serves as an intrinsic radioactive source, enabling a source-in-detector configuration with high detection efficiency and minimal energy loss for low-energy electrons from beta-decay.
We investigated both Rb doped and Tl co-doped CsI crystals and characterized their scintillation properties, including light yield, energy resolution, and non-linear response. We report distinct scintillation characteristics for the CsI:Rb and CsI:Tl,Rb crystals, with light yields of $1.38\pm0.01$ and $4.73\pm0.13$\,PE/keV, respectively.
Using the measured $^{87}$Rb beta-spectrum, we search for a keV-scale sterile neutrino admixture through the characteristic kink-like distortion induced by a heavy neutrino mass eigenstate. This study provides a basis for future sterile neutrino searches using rubidium doped CsI crystal.


\end{abstract}

\begin{IEEEkeywords}
Rubidium-87, CsI, Crystal scintillator, beta-spectroscopy, beta-decay, sterile neutrino, dark matter
\end{IEEEkeywords}

\section{Introduction}
\label{sec:introduction}
Rubidium-87 undergoes a third-forbidden non-unique $\beta^-$ transition to the ground state of $^{87}$Sr. The current evaluated $\text{Q}_\beta$ value is 282.275(6)\,keV~\cite{mougeot2025,wang2021}, and the half-life $T_{1/2}$ is 4.9650(40)$\times10^{10}$\,yr~\cite{villa2015, mougeot2025}. Because several nuclear matrix elements (NMEs) contribute to a non-unique forbidden transition~\cite{suhonen2017, ejiri2019}, precise measurements and calculations of the $^{87}$Rb spectrum have therefore been used for decades to constrain the shape factor $C(W)$~\cite{kossert2003, graucarles2006, szyb, beard1961, ruettenauer1973}
and nuclear structure descriptions~\cite{macgregor1954, flynn1959, szyb, sastry1969, beard1961, kossert2003,graucarles2006}. 
A precision measurement of the $^{87}$Rb beta-spectrum can provide experimental constraints on NME calculations and the effective value of the axial-vector coupling constant $g_A$, contributing to the understanding of forbidden beta-decay. 

Beyond nuclear studies, beta-spectrum data measured with rubidium-doped CsI crystal can probe physics beyond the Standard Model (SM). In particular, it can be used to search for sterile neutrinos, a candidate for dark matter, through the mixing of the electron neutrino with a sterile state, which would produce a second kinematic endpoint and a kink-like distortion in the spectrum~\cite{morita1963, ejiri2019, dasgupta2021a, friedrich2021, boyarsky2019}.

In this work, we investigate the incorporation of a large nominal concentration of rubidium into a CsI scintillator. At such a high loading, the principal material questions are whether a coherent crystal can be grown and whether the resulting material provides a stable and reproducible scintillation response suitable for beta-spectroscopy. We therefore compare a CsI crystal grown with a nominal RbI loading of 10\,mol\% with a second crystal in which Tl is introduced as a co-dopant.

\section{Experimental Setup}
\label{sec:setup}

\subsection{Crystal growing}
\label{subsec:crystalgrowing}

\begin{figure}[b]
\centerline{\includegraphics[width=0.7\linewidth]{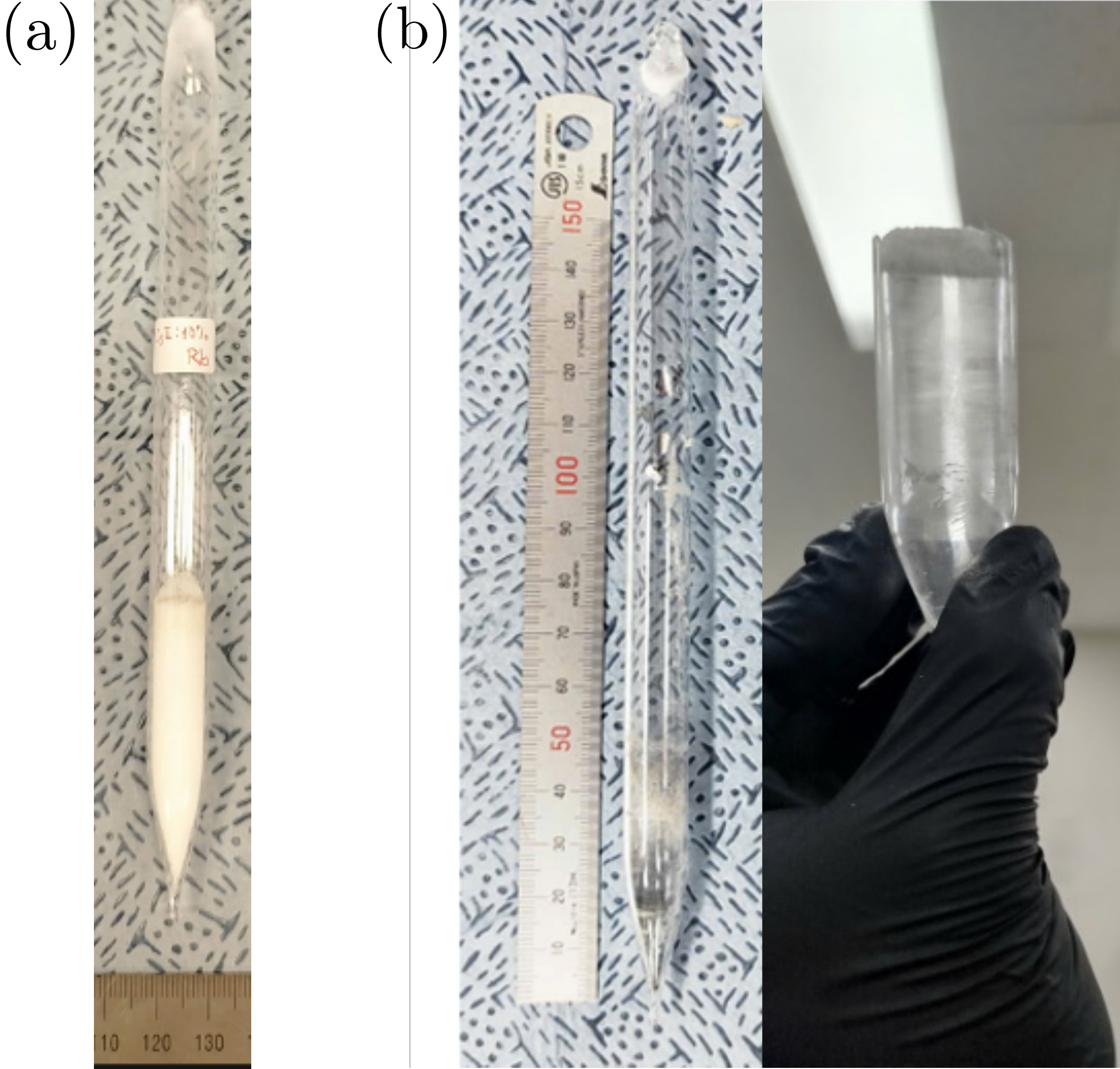}}
\caption{(a) Crystal powder in preparation. (b) Grown CsI:Rb crystal with Bridgman method.}
\label{fig:growing}
\end{figure}

\begin{figure*}[t]
    \centering
    \includegraphics[width=0.8\textwidth]{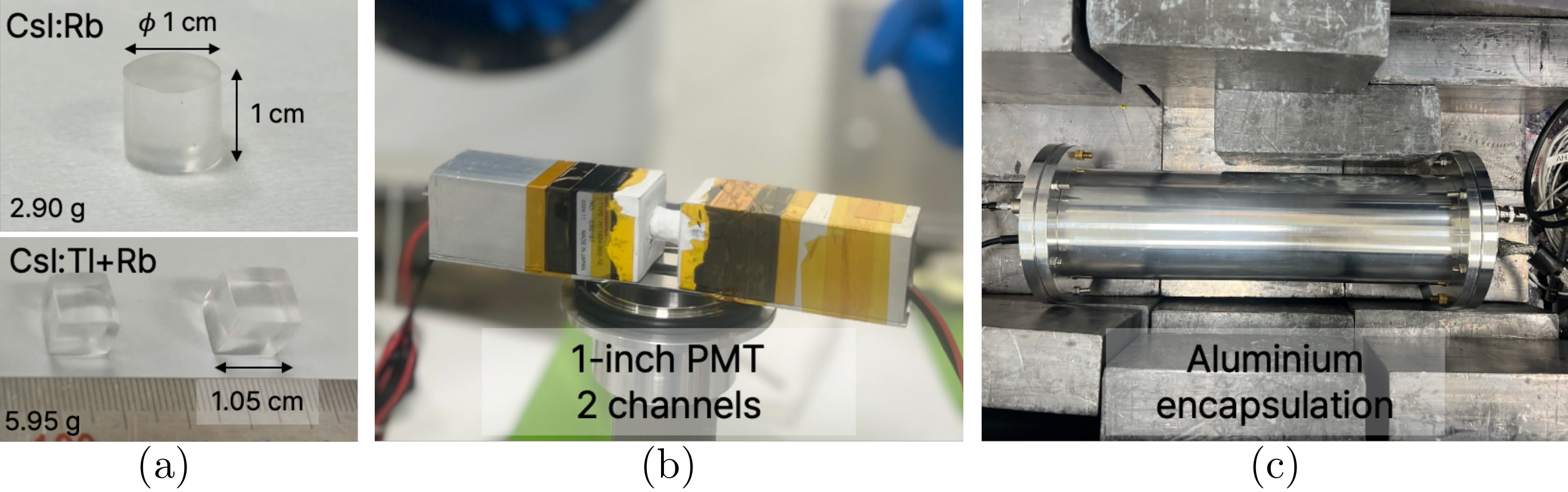}
    \centering
    \caption{Detector assembly process for these  measurements. (a) 10\,mol\% RbI-doped crystals with different geometry. A cylindrical shape for CsI:Rb, which has a 10\,mm $\phi\times10$\,mm with 2.90\,g. A CsI:Tl,Rb is a 5.95\,g of cubic crystal with 1.05\,cm$\times$1.05\,cm$\times$1.05\,cm dimension. (b) Two 1-inch Hamamatsu PMTs were coupled at the both end sides. For the effective photon collection, layers of soft polytetrafluoroethylene (PTFE) sheets are wrapped around the exposed area of PMTs and surface of the crystals. (c) To shield against light and moisture, the detector is encased in an aluminum housing.}
    \label{fig:setup}
\end{figure*}

Purified CsI powder from the KIMS dark matter direct-detection experiment~\cite{lee2005} was used as the base material. RbI powder with a purity of 3N (Sigma-Aldrich) was added to achieve a nominal RbI loading of 10\,mol\%.
For the CsI:Tl,Rb crystal, TlI powder supplied by the Center for Underground Physics at the Institute for Basic Science was additionally introduced at a concentration of 0.1\,mol\%. All weighing, mixing, and loading were performed in a nitrogen-purged glovebox and the mixture is shown in Fig.~\ref{fig:growing}(a). In the glovebox, the water concentration was maintained below approximately 300--400\,ppm at 25\,$^{\circ}$C. Quartz ampoules were sequentially rinsed three times with 99.9\,\% ethanol, three times with deionized water, and then dried for 2--3\,d. After the powder mixture was loaded, each ampoule was evacuated to below $10^{-6}$\,Torr and heated at 120\,$^\circ$C for 24\,h for further drying before being sealed using a propane torch.

The sealed ampoule was placed in a two-zone vertical Bridgman furnace at Kyungpook National University. The two-zone temperature configuration promoted melt convection before directional solidification. Both zones were initially heated to approximately 700\,$^\circ$C to fully melt the charge, with mixing promoted by melt convection. The ampoule was then positioned near the inter-zone gap and the zone temperatures were adjusted to produce a gradient of approximately 10\,$^\circ$C/cm at the solid-liquid interface. For the nominal 10\,mol\% RbI material, the ampoule was translated downward at 0.6\,mm/h. After solidification was complete, both furnace zones were cooled to room temperature at 6\,$^\circ$C/h. Visual inspection showed coherent ingot formation as shown in Fig.~\ref{fig:growing}(b).

\begin{figure}[htb!]
\centerline{\includegraphics[width=3.3in]{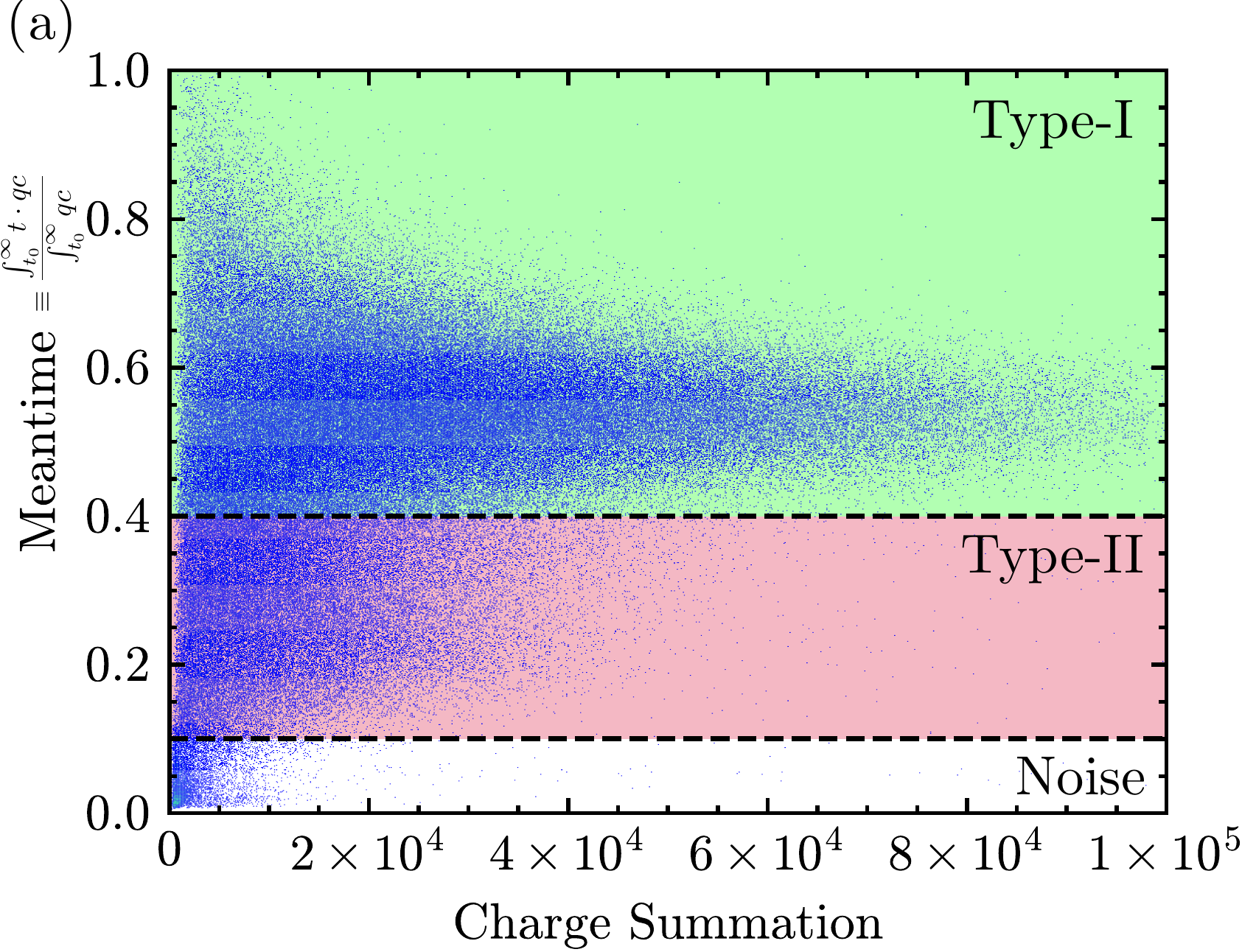}}
\centerline{\includegraphics[width=3.1in]{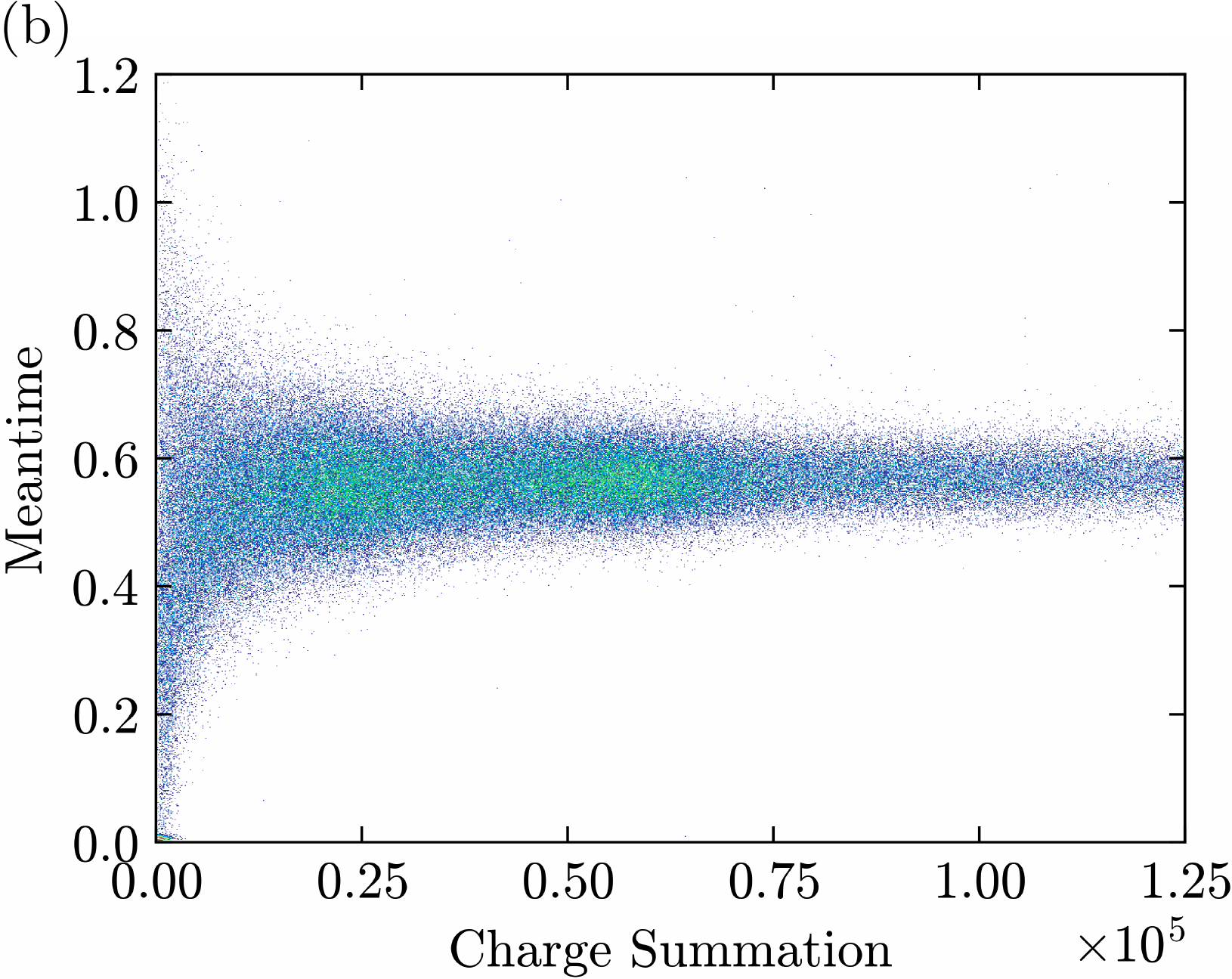}}
\caption{A meantime parameter versus charge summation from the waveform in $^{133}$Ba data. (a) CsI:Rb crystal shows two branches, while (b) CsI:Tl,Rb crystal shows one population branch.}
\label{fig:twobranches}
\end{figure}

\begin{figure}[htb]
\centerline{\includegraphics[width=3.1in]{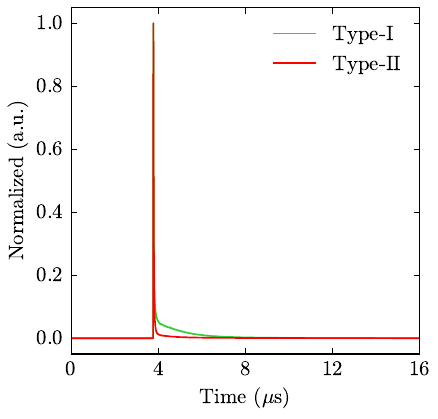}}
\caption{Averaged waveform from the CsI:Rb crystal. Type-I and Type-II show different decay times for the scintillation events.}
\label{fig:waveform}
\end{figure}

\subsection{Detector assembly and data acquisition}
\label{subsec:detector}
The CsI:Rb crystal has a cylindrical shape, 10\,mm in diameter and 10\,mm long, with a mass of 2.90\,g, while the CsI:Tl,Rb crystal is a 5.95\,g cube, as shown in Fig.~\ref{fig:setup}(a). Both crystals were optically coupled to Hamamatsu H11934-300-10 photomultiplier tube (PMT) modules with integrated readout electronics (Fig.~\ref{fig:setup}(b)). The H11934-300-10 is an extended green bialkali PMT, which provides a higher quantum efficiency than conventional bialkali PMTs in the emission wavelength range of CsI crystals, making it well suited for efficient collection of the scintillation light from both CsI:Rb and CsI:Tl,Rb. To protect the crystals from light and moisture, the assembled detector was enclosed in an aluminium housing (Fig.~\ref{fig:setup}(c)). The scintillation signals were amplified by a factor of 30 using an external amplifier and digitized with a 500\,MHz, 12-bit flash-analogue-to-digital converter (FADC). An event was triggered when a coincident pulse was observed in both PMTs within a 200\,ns window, and the waveform was recorded over a 16\,\textmu s acquisition window. 

For energy calibration, data were acquired with $^{241}$Am, $^{133}$Ba, $^{57}$Co, and $^{137}$Cs sources. Because the crystal contains intrinsic $^{87}$Rb activity, a source-off background spectrum was subtracted from each source-run spectrum prior to peak extraction for calibration.
The $^{87}$Rb beta-spectrum measurement was additionally performed with the encapsulated detector immersed in a linear-alkylbenzene-based (LAB) liquid scintillator (LS), which served as a shield for external backgrounds.

\section{Scintillation Response of CsI:Rb and CsI:Tl,Rb}
\label{sec:scintillation}
\subsection{Pulse shape characteristics and light yield}
\label{subsec:lightyield}


The most striking difference between the two crystals appears in the meantime plotted against integrated waveform charge, as shown in Fig.~\ref{fig:twobranches}, where the meantime is defined as the charge-weighted average time of the waveform,
\begin{equation}
\text{Meantime}\equiv \frac{\displaystyle\int_{t_0}^{t_0+3\,\text{$\mu$s}} t \cdot q_c(t)\,dt}{\displaystyle\int_{t_0}^{t_0+3\,\text{$\mu$s}} q_c(t)\,dt},
\label{eq:meantime}
\end{equation}
where $t_0$ is the start time of the waveform and $q_c(t)$ is the charge collected at time $t$. 
Rb doping of CsI is expected to modestly increase the light yield and to produce a somewhat longer decay time relative to pure CsI. However, the meantime distribution of CsI:Rb separates into two distinct branches (Fig.~\ref{fig:twobranches}(a)): one with larger charge and longer meantime, consistent with the expected effect of Rb doping (Type-I), and a second with smaller charge and shorter meantime, resembling the scintillation characteristics of pure CsI (Type-II). Averaged waveforms formed from the two branches indeed show substantially different decay times, as depicted in Fig.~\ref{fig:waveform}. In~\cite{gang2003}, the scintillation characteristics of a rubidium-doped CsI crystal were described; however, such two kinds of scintillation processes have not been previously reported. The presence of two distinct scintillation populations in CsI:Rb suggests that the Rb dopant may not be uniformly distributed within the crystal, or may not fully function as an effective activator, allowing pure CsI-like scintillation to persist alongside the Rb-doped component; further study is needed to confirm this hypothesis. For subsequent analysis, events in the longer decay time branch, consistent with the expected Rb-doping effect, were selected using a cut on the meantime.

In contrast, the CsI:Tl,Rb crystal exhibits a single population, without the two distinct branches observed in CsI:Rb, as shown in Fig.~\ref{fig:twobranches}(b). This behavior, consistent with that previously reported for CsI:Tl crystals~\cite{Kim:2003ms,Lee:2007iq,schotanus1990,gwin1963,gwin1963a}, indicates that Tl co-doping effectively suppresses the pure CsI-like scintillation component, yielding a single, well-defined population.

\begin{figure}[t]
\centerline{\includegraphics[width=3.4in]{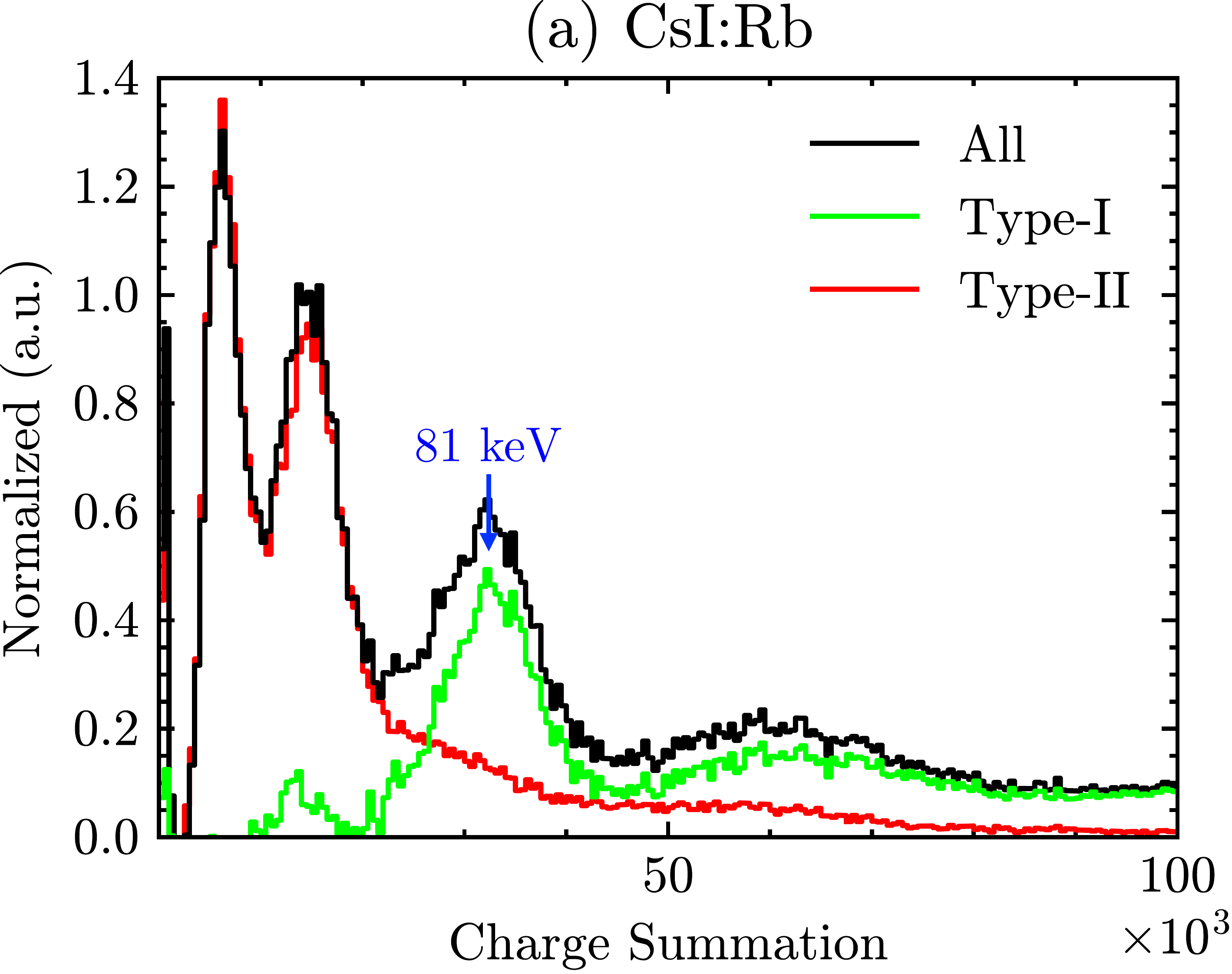}}
\vspace{2mm}
\centerline{\includegraphics[width=3.3in]{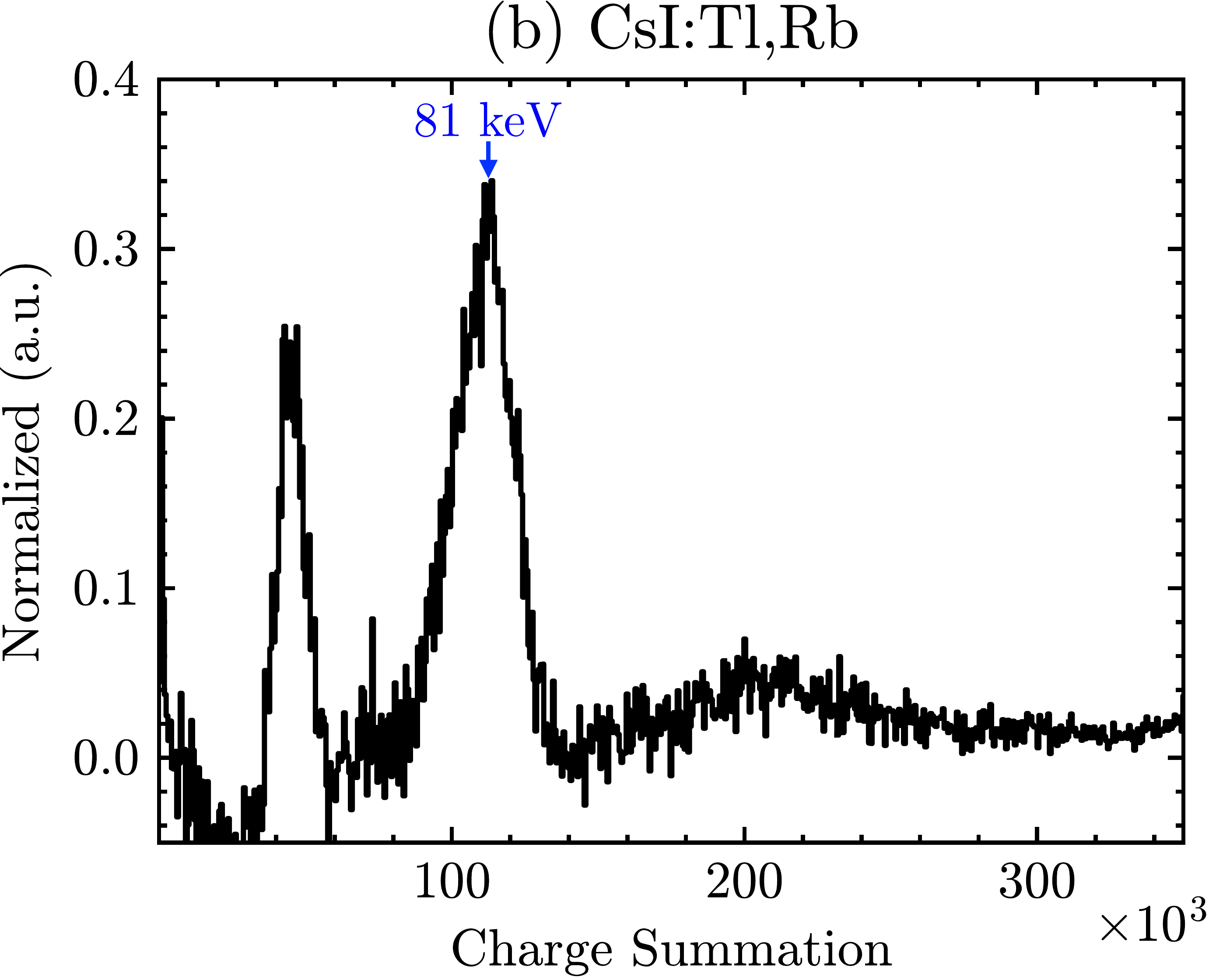}}
\caption{Spectrum of $^{133}$Ba source is drawn with black line. To obtain the light yield of the crystal, 81\,keV of $\gamma$-ray peak was leveraged. (a) For CsI:Rb crystal, Type-I events were selected through the meantime parameter cut, and its energy peak is drawn with green line. Type-II events are drawn with red line. (b) Spectrum for CsI:Tl,Rb crystal. The 30.85\,keV peak from $^{133}$Ba was excluded from the light yield evaluation because the trigger efficiency was not fully characterized in this low-energy region.}
\label{fig:ba133}
\end{figure}

The light yield was evaluated using the 81\,keV $\gamma$-ray peak from $^{133}$Ba, as shown in Fig.~\ref{fig:ba133}. For CsI:Rb as shown in Fig.~\ref{fig:ba133}(a), the green spectrum corresponds to events in the longer meantime branch, consistent with the expected effect of Rb doping, while the red spectrum corresponds to events in the shorter meantime branch, in which the Rb-doping effect appears largely absent; the light yield of CsI:Rb was evaluated using the green spectrum, corresponding to the longer meantime branch selected as described above. For CsI:Tl,Rb, consistent with the single population discussed above, the 81\,keV peak appears as a single, well-defined feature in Fig.~\ref{fig:ba133}(b). The light yield of CsI:Rb was measured to be $1.38\pm0.01$\,PE/keV, while that of CsI:Tl,Rb was $4.73\pm0.13$\,PE/keV, which is a factor of 3.4 higher than that of CsI:Rb. The result demonstrates a substantially larger collected light yield after Tl co-doping for the present detector configurations. However, because of the difference in crystal mass and geometry between the two samples, a future precise comparison should also quantify photon collection and optical coupling systematics.

Given the higher and more stable light yield achieved with Tl co-doping, the CsI:Tl,Rb crystal was adopted for the $^{87}$Rb beta-spectrum measurement, and its energy resolution was evaluated using the 59.54\,keV line from $^{241}$Am, the 80.99\,keV line from $^{133}$Ba, the 122.06\,keV line from $^{57}$Co, and the 661.66\,keV line from $^{137}$Cs. The energy resolution was parameterized as
\begin{equation}
\frac{\sigma}{E} = \frac{p_{0}}{\sqrt{E}} + p_{1},
\label{eq:resolution}
\end{equation}
where $p_{0}$ and $p_{1}$ are free parameters. A fit to the measured resolution values yielded $p_{0} = 60.19$ and $p_{1} = 2.42$, with $\chi^2$ per number of degrees of freedom (ndf) of 1.05.



\subsection{Nonproportionality of gamma response in CsI:Tl,Rb}
\label{subsec:nonproportionality}
\begin{figure}[htb]
\centerline{\includegraphics[width=3.2in]{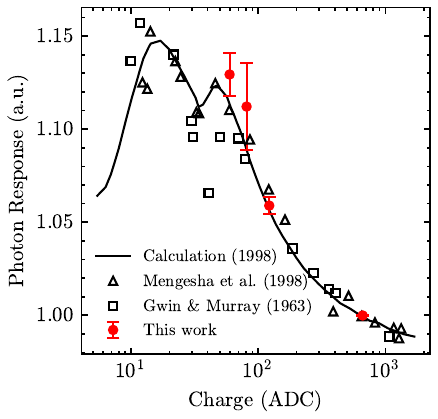}}
\caption{Gamma response of the calculation and measurement value of CsI:Tl~\cite{mengesha1998,gwin1963}, and CsI:Tl,Rb measured in this work. Data points are all normalized at 661.66\,keV.}
\label{fig:gammaresponse}
\end{figure}

The light yield of inorganic scintillators is known to exhibit a nonproportional dependence on the deposited energy~\cite{mengesha1998,gwin1963}, which directly affects the accuracy of the energy calibration, particularly at low-energies. To investigate this effect for the CsI:Tl,Rb crystal, the relative photon response was evaluated from the detected light yield at each $\gamma$-ray energy and normalized to the response at 661.66\,keV.
Fig.~\ref{fig:gammaresponse} compares the present CsI:Tl,Rb data with calculated photon response curves and measured $\gamma$-ray data reported for CsI:Tl in~\cite{mengesha1998, gwin1963}.
Across the measured energy range, CsI:Tl,Rb broadly follows the characteristic CsI:Tl response, with a slight increase in nonproportionality at few-keV energies. The comparison, therefore, suggests that the Rb doped CsI:Tl crystal has a similar $\gamma$ and electron response to the conventional CsI:Tl. However, the lack of theoretical calculations below 10\,keV for CsI:Tl makes it difficult to draw a firm conclusion about the low-energy response of CsI:Tl,Rb. Thus, a quadratic function was fitted to the present data to provide a simple parameterization of the nonproportionality for the beta-spectrum analysis. This quadratic function was adopted as the energy calibration for the subsequent $^{87}$Rb beta-spectrum analysis, in order to account for the nonproportional response of the detector.
The systematic uncertainty from the nonproportionality curve is under investigation and will be included in the future analysis with larger statistics.

\section{Beta-Spectroscopy of $^{87}$Rb}
\label{sec:spectroscopy}

\subsection{Shape factor}
\label{subsec:shapefactor}
For a third-forbidden non-unique decay, a commonly used spectral shape $C(W)$ is parameterized, containing multiple momentum-dependent terms,
\begin{equation}
C(W) = q^4 + ap^2q^2 + bp^4,
\label{eq:cw}
\end{equation}
where $q$ and $p$ are the neutrino and electron momenta, respectively, and $a$ and $b$ are shape parameters~\cite{graucarles2006, szyb, ruettenauer1973, beard1961}, which cannot be derived fully by theoretical calculations~\cite{szyb}. Extracting $a$ and $b$ from the measured spectrum requires a precise understanding of the detector response, including energy resolution, nonproportionality, and background contributions. This provides sensitivity to the effective axial-vector coupling constant $g_A$ and the combinations of NME that contribute to this non-unique transition~\cite{suhonen2017, ejiri2019}.

\begin{table}[b]
\caption{Comparison of the shape factor $C(W)$ coefficients for the third-forbidden non-unique beta-decay of $^{87}$Rb}
\label{tab:cw}
\setlength{\tabcolsep}{3pt}
\renewcommand{\arraystretch}{1.2}
\begin{tabular}{|p{80pt} p{67.5pt} p{67.5pt}|}
\hline
Measurement&
a&
b\\
\hline
K.~Kossert \textit{et al.}~\cite{kossert2003}&
$0.354$&
$0.003$\\
A.~Grau \textit{et al.}~\cite{graucarles2006}&
$0.305 \pm 0.009$&
$0.011 \pm 0.008$\\
P.~Belli \textit{et al.}~\cite{belli2026}&
$0.3607 \pm 0.001$&
$0.00620 \pm 0.00003$\\
This work&
$0.3481 \pm 0.0018$&
$0.00616 \pm 0.00007$\\
\hline
\multicolumn{3}{p{251pt}}{The coefficients $a$ and $b$ are defined through $C(W) = q^4 + ap^2q^2 + bp^4$, where $p$ and $q$ denote the electron and neutrino momenta, respectively. The uncertainties for this work are derived from statistical error. The best-fit values obtained in this work were fixed in the sterile neutrino search.}
\end{tabular}
\end{table}

\begin{figure}[t]
\centerline{\includegraphics[width=3.6in]{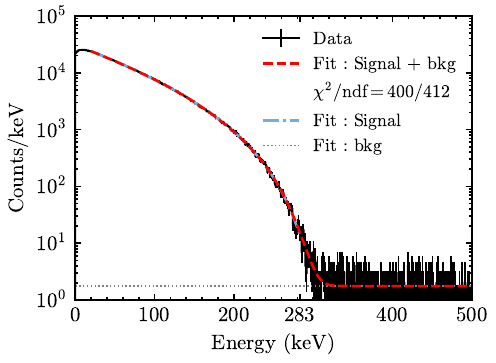}}
\caption{Measured beta-spectrum of $^{87}$Rb obtained with the CsI:Tl,Rb crystal over a 5\,hr exposure (Data). The spectrum is fitted with a function including the theoretical beta-spectrum (Signal only), a constant background component (bkg), and their sum (Signal + bkg).}
\label{fig:betaspectrum}
\end{figure}

A 5\,hr measurement of the CsI:Tl,Rb crystal was performed within the LAB-based LS setup described in section~\ref{subsec:detector}.
The observed spectrum is described by a function that includes the detector response, Fermi function, and a constant background component, and the fit was conducted from 20\,keV to 500\,keV. We obtained $a = 0.3481 \pm 0.0018$ and $b = 0.00616 \pm 0.00007$ with $\chi^2$/ndf = 400/412. The extracted shape parameters could be compared with other experimental results in Table~\ref{tab:cw} and theoretical calculations~\cite{szyb,ruettenauer1973,kossert2003,graucarles2006, belli2026}. 
The fit result can also be seen as the red dashed curve in Fig.~\ref{fig:betaspectrum}. While the region above end point energy $\text{E}_0$ constrains the background level, the blue curve represents a signal only fit restricted to the range from 20\,keV to $\text{E}_0$. 

\subsection{Sterile neutrino search}
\label{subsec:sterile}
In beta-decay, an admixture of a heavier mass eigenstate $\nu_4$ in the electron-flavor neutrino produces an additional spectral component terminating at $\text{E}_0-m_4$, resulting in a kink distortion whose magnitude is governed by the mixing parameter $|\text{U}_{e4}|^2$~\cite{boyarsky2019}. In the present analysis, the coefficients of the shape factor in Eq.~\eqref{eq:cw} were fixed to the values that we obtained in~\cref{subsec:shapefactor}. 
For a representative sterile neutrino mass of $m_4 = 106.3$\,keV/$c^2$, a statistical-only upper limit of $|\text{U}_{e4}|^2 < 6.8 \times 10^{-3}$ was obtained at 95\% confidence level (C.L.). Systematic uncertainties arising from the energy calibration and nonproportionality, energy resolution, background modeling, endpoint energy, and the shape factor coefficients remain to be evaluated and incorporated into both the shape factor and the limit.


\section{Conclusion}
\label{sec:conclusion}
CsI crystals with a high rubidium doping concentration of 10\,mol\% were grown using the Bridgman method and investigated for source-in-detector beta-spectroscopy of $^{87}$Rb.
The CsI:Rb crystal exhibited two distinct scintillation populations, whereas the addition of 0.1\,mol\% Tl doped, CsI:Tl,Rb crystal, resulted in a single population. The energy calibration was conducted with external gamma sources, and the light yield was $1.38\pm0.01$\,PE/keV for CsI:Rb and $4.73\pm0.13$\,PE/keV for CsI:Tl,Rb. 
Since the two relaxation paths in the CsI:Rb crystal require further study, the energy resolution and nonproportionality were characterized for the CsI:Tl,Rb scintillation detector, whose relative photon response exhibited a nonproportionality trend broadly consistent with that of conventional CsI:Tl crystals. These results indicate that Tl co-doping provides a more stable and higher light yield scintillation response for $^{87}$Rb beta-spectroscopy.

Five hours of data were collected with the detector immersed in LAB-based LS to suppress the external background level. A preliminary fit incorporating the theoretical beta-spectrum, detector response, and a constant background component yielded the shape factor parameters $a = 0.3481 \pm 0.0018$ and $b = 0.00616 \pm 0.00007$ with statistical-only uncertainties. 
Using these best-fit shape factor coefficients as fixed inputs, a search for sterile neutrino admixture was also performed. For a representative sterile neutrino mass of $m_4 = 106.3$\,keV/$c^2$, the statistical-only upper limit was $|\text{U}_{e4}|^2<6.8\times10^{-3}$ at 95\% C.L., while systematic uncertainties remain to be evaluated. 
A higher statistics measurement is planned at Yemilab~\cite{kim2024}, where the underground operation and improved shielding are expected to substantially reduce cosmic ray induced and external backgrounds.
With such increased exposure and a complete treatment of systematic errors, this approach could provide precise measurements on the relevant NME and the effective axial-vector coupling constant $g_A$, as well as sensitivity to sterile neutrino admixtures.


\bibliographystyle{IEEEtran}
\bibliography{Ref_csirb}

\end{document}